\documentclass{article}
\usepackage{setspace, emlines2}
\begin{document}
\large \pagestyle{plain}

\centerline{\LARGE The Quantum Liar Experiment in Cramer's Transactional Interpretation}

\vskip 1.5cm \centerline{Ruth E. Kastner}

\centerline{University of Maryland, College Park}
 \centerline{Version of Dec. 28, 2010} \vskip .5cm

ABSTRACT. Cramer's Transactional Interpretation (TI) is applied to the ``Quantum Liar Experiment'' (QLE). It is shown how some apparently paradoxical features can be explained naturally, albeit nonlocally (since TI is an explicitly nonlocal interpretation, at least from the vantage point of ordinary spacetime). At the same time, it is proposed that in order to preserve the elegance and economy of the interpretation, it may be necessary to consider offer and confirmation waves as propagating in a ``higher space'' of possibilities.
\vskip .3cm

\vskip .5 cm
\onehalfspacing

\newcommand{\ketzu}{|z\!\uparrow\rangle\;}
\newcommand{\ketzd}{|z\!\downarrow\rangle\;}
\newcommand{\brazu}{\langle z \uparrow|}
\newcommand{\brazd}{\langle z\downarrow|}
\newcommand{\brayu}{\langle y \uparrow|}
\newcommand{\brayd}{\langle y \downarrow|}
\newcommand{\ketd}{|d\rangle\;}
\newcommand{\ketc}{|c\rangle\;}
\newcommand{\ketu}{|u\rangle\;}
\newcommand{\ketv}{|v\rangle\;}
\newcommand{\bsu}{|++\rangle\;}
\newcommand{\bsd}{|--\rangle\;}
\newcommand{\ud}{|+-\rangle\;}
\newcommand{\du}{|-+\rangle\;}
\newcommand{\gr}{|0\rangle\;}
\newcommand{\ex}{|1\rangle\;}
\newcommand{\udex}{|+-,1\rangle\;}
\newcommand{\dugr}{|-+,0\rangle\;}
\vskip .5cm

{\bf 1. Introduction}\vskip .5cm

The Quantum Liar Experiment is an ingenious gedanken experiment first suggested by Elitzur, Dolev and Zeilinger [2002], based on pioneering work by Elitzur and Vaidman [1993] on ``interaction free measurements'' (IFM) and ensuing work by Hardy [1992]. Much has been written already about such experiments,\footnote{\normalsize Some of the many papers on IFM will be discussed and referenced in what follows.} and deservedly so, since they exhibit very clearly the nonclassical nature of quantum events. This paper will consider the QLE and related experiments in the light of the Transactional Interpretation (TI) of John G. Cramer, first proposed in 1980 and presented in a comprehensive manner in his [1986]. It will further
suggest a new variant of TI, called ``Possibilist TI" or PTI.

It is this author's view that TI is a seriously underappreciated interpretation that offers the most natural and elegant approach to the many conceptual challenges of quantum theory, albeit at the price of a profound paradigm change. Key features of the new paradigm are (1) time-reversed influences and (2) physical processes that operate at a level of possibility rather than at the level of actualized reality. Yet TI merits due consideration since the former is implicit in the quantum  formalism. Moreover, Feynman's elegant and powerful sum-over-histories approach explicitly suggests the latter (i.e., that a lot of important activity seem to go on ``behind the scenes'' in any quantum process). There is ample evidence of a need for such a paradigm change when we look at the outright contradictions and inconsistencies that arise in the context of many recent proposed experiments that make the EPR experiment seem tame by comparison.

We should first note that Cramer's analysis of IFM [Cramer, 2006] emphasizes that TI is an explicitly nonlocal and atemporal interpretation, and that it makes no attempt to give a ``local'' account on the level of determinate particles (because according to TI, these are not fundamental anyway). TI is considered causal, however, to the extent that the wavelike entities represented by quantum states are seen as functioning dynamically in physical interactions with each other and with the experimental apparatus\footnote{\normalsize{To be precise, a full account would consider the experimental apparatus as part of the transactional interplay. Apparently persistent macroscopic objects have the same ontological status as microscopic objects in this interpretation. Their persistence arises from very frequent and overwhelmingly probable transactions which serve to constrain the possible transactions available to the microscopic systems under study.}}  in the form of retarded (normal time sense)``offer waves'' (OW) and advanced (time-reversed) ``confirmation waves'' (CW), identified  as $|\psi\rangle$ and $\langle\psi |$, respectively. For example, Cramer's account of Elitzur-Vaidman's [1993] interaction-free detection of a bomb on one of two arms of a Mach-Zehnder Interferometer [MZI] involves offer waves going along both paths even though one might be blocked. The information obtained in the experiment is attributed specifically to this nonlocal character of a wave associated with a `potential' particle but yet not giving rise to a detected particle in that location. I should also note here that in his (1986), Cramer applied TI to the Hanbury-Twiss effect which EDZ utilize in a version of the QLE in their (2002). Cramer's interpretation of the H-T effect will be addressed in connection with this version of the QLE in part 4.

\vskip .5cm
{\bf 2. A quantum ``bomb''.} \vskip .5cm

Before dealing with the QLE, we first consider a simpler setup: Hardy's twist on the original Elitzur-Vaidman IFM, in which a bomb or other obstruction is placed inside one arm of an MZI. Recall that in the E-V version, the MZI is tuned so that one of the detectors, which we will call D, will never activate unless something is obstructing one of the paths. In Hardy's version, the bomb or other macroscopic object is replaced by a quantum system: a spin one-half atom. The atom is prepared in a state of ``up along x'' which is then subject to a magnetic field gradient along the z direction and spatially separated so that it could be found in either of two boxes, one of which (``up along z'' or $|\:z\uparrow\rangle$) is carefully placed in one path of the MZI. (refer to Figure 1.)\vskip .5cm

\unitlength 1mm 
\linethickness{0.4pt}
\ifx\plotpoint\undefined\newsavebox{\plotpoint}\fi 
\begin{picture}(129.5,107)(0,0)
\put(40.25,20.75){\framebox(53.25,53)[cc]{}}
\put(93.75,73.75){\line(1,0){27}}
\put(93.25,73.75){\line(0,1){24.75}}
\put(93.125,101.375){\oval(4.75,4.75)[t]}
\put(123.375,73.5){\oval(4.75,5)[r]}
\multiput(90.25,17.25)(.033653846,.037259615){208}{\line(0,1){.037259615}}
\multiput(36,70.25)(.038372093,.03372093){215}{\line(1,0){.038372093}}
\multiput(88.18,69.93)(.0411765,.0323529){17}{\line(1,0){.0411765}}
\multiput(89.58,71.03)(.0411765,.0323529){17}{\line(1,0){.0411765}}
\multiput(90.98,72.13)(.0411765,.0323529){17}{\line(1,0){.0411765}}
\multiput(92.38,73.23)(.0411765,.0323529){17}{\line(1,0){.0411765}}
\multiput(93.78,74.33)(.0411765,.0323529){17}{\line(1,0){.0411765}}
\multiput(95.18,75.43)(.0411765,.0323529){17}{\line(1,0){.0411765}}
\multiput(96.58,76.53)(.0411765,.0323529){17}{\line(1,0){.0411765}}
\multiput(97.98,77.63)(.0411765,.0323529){17}{\line(1,0){.0411765}}
\multiput(34.68,15.93)(.0378289,.0320724){19}{\line(1,0){.0378289}}
\multiput(36.117,17.148)(.0378289,.0320724){19}{\line(1,0){.0378289}}
\multiput(37.555,18.367)(.0378289,.0320724){19}{\line(1,0){.0378289}}
\multiput(38.992,19.586)(.0378289,.0320724){19}{\line(1,0){.0378289}}
\multiput(40.43,20.805)(.0378289,.0320724){19}{\line(1,0){.0378289}}
\multiput(41.867,22.023)(.0378289,.0320724){19}{\line(1,0){.0378289}}
\multiput(43.305,23.242)(.0378289,.0320724){19}{\line(1,0){.0378289}}
\multiput(44.742,24.461)(.0378289,.0320724){19}{\line(1,0){.0378289}}
\put(62.75,17){\dashbox{1}(6.75,7)[cc]{}}
\put(62.75,5.75){\dashbox{1}(6.75,7)[cc]{}}
\put(51.5,22.75){\makebox(0,0)[cc]{$v$}}
\put(44.25,33.75){\makebox(0,0)[cc]{$u$}}
\put(97.25,84.5){\makebox(0,0)[cc]{$d$}}
\put(104.75,70.75){\makebox(0,0)[cc]{$c$}}
\put(93,107){\makebox(0,0)[cc]{D}}
\put(129.5,73.75){\makebox(0,0)[cc]{C}}
\put(40.25,21){\line(-1,0){15}}
\put(22.25,21){\makebox(0,0)[cc]{L}}
\put(66.5,20.25){\makebox(0,0)[cc]{$\uparrow$}}
\put(66.5,8.75){\makebox(0,0)[cc]{$\downarrow$}}
\put(40.75,15.5){\makebox(0,0)[cc]{S1}}
\put(89.5,77.5){\makebox(0,0)[cc]{S2}}
\put(38.25,76.25){\makebox(0,0)[cc]{B}}
\put(96,19){\makebox(0,0)[cc]{A}}
\end{picture}

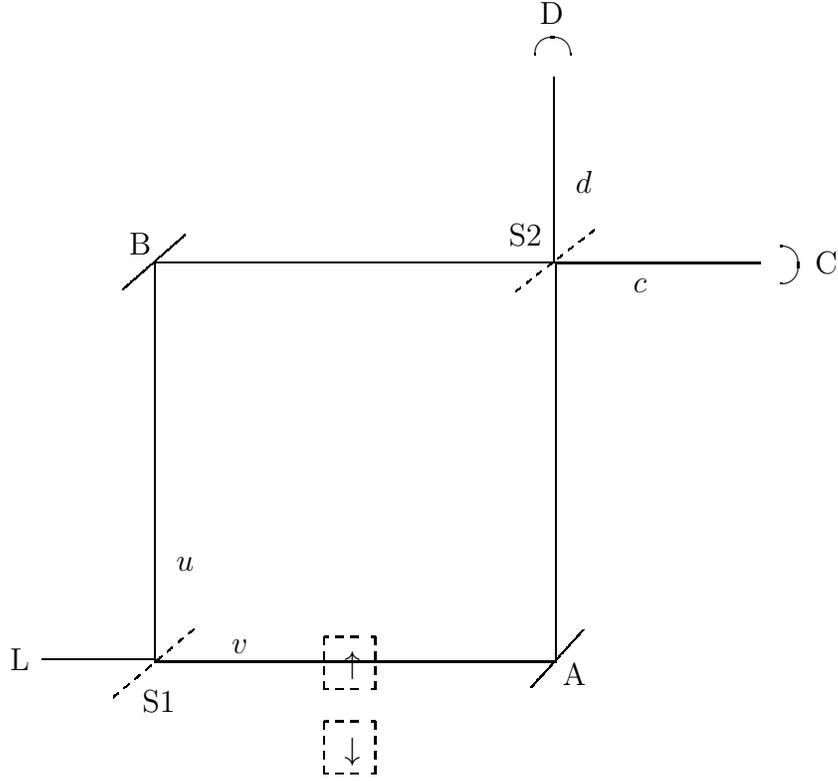
 \normalsize\centerline{Figure 1. Hardy's version of the Elitzur-Vaidman Interaction Free Measurement with an atom replacing the bomb.}
 \vskip .5cm
 \large

 As noted in Hardy's discussion and by Elitzur, Dolev and Zeilinger, the surprising feature of this experiment is that when detector D is activated, the atom is determined to be in the box intersecting path $v$ in a well-defined spin state $|\:z\uparrow\rangle$, yet seemingly the photon did not interact with it since the latter was detected at D. How is this possible? Hardy's discussion is centered around the idea of ``empty waves,'' i.e., Bohmian guiding waves in which the particle is clearly absent yet the wave appears to have real effects. It is not our purpose here to address the Bohmian ``empty wave'' picture but to show that TI gives a very natural and revealing account of this experiment.

The atom is understood to be in its ground state $\gr$ unless otherwise specified. The atom's excited state is denoted as $\ex$. The state of the combined system of \{photon, atom\} starts out as:

 $$ {|\:\Psi\rangle}_i = |\:s\rangle \otimes {1 \over {\sqrt 2}} [{|\: z\!\uparrow\rangle + |\: z\!\downarrow\rangle}].   \eqno(1)$$

After passing through the first beam splitter S1, the photon's state becomes ${1\over {\sqrt 2}} (i |\:u\rangle + |\:v\rangle)$,
so at this point the total system state is:

$$ {|\:\Psi\rangle} = {1\over 2} ( i |\:u\rangle + |\:v\rangle) \otimes  [{|\: z\!\uparrow\rangle + |\: z\!\downarrow\rangle}]$$

$$ =  \frac {1}{2} [i \ketu\! | z\!\uparrow\rangle  + \ketv\! |\: z\!\uparrow\rangle +
i |\:u\rangle | z\!\downarrow\rangle +  |\:v\rangle |\: z\!\downarrow\rangle].\eqno(1a)$$

Now, under TI this is considered to be an offer wave (OW). The second term involves a potential transaction corresponding to
the photon being found on path $v$ and the atom occupying the intersecting box. Under the idealized assumptions
of the experiment, the actualization of this transaction will result in absorption of the photon by the atom. But this
can only occur in the presence of an atomic confirmation wave in the excited state $\langle z\!\uparrow, 1|$. So TI needs
to allow for a hierarchy of transactions in which those with a shorter spacetime interval have ontological
priority over other transactions with a longer spacetime interval. Thus transactions involving photon detections at C or D are contingent on the failure of the transaction involving absorption by the atom. Cramer discussed this
 feature of TI in a powerpoint presentation (2005).\footnote{\normalsize The initial observation of this type of challenge for TI was made by Tim Maudlin (2002) and, in addition to Cramer's proposed solution of a hierarchy (2005), was addressed in Kastner (2006) as well in Berkovitz (2002). In his (2002), Maudlin also raised some other conceptual challenges for TI, which I believe are at least partially resolved by the approach proposed herein.}

If the absorption transaction does not occur,
 the remaining photon OW proceeds through the final beam splitter S2 and the system state evolves at the detector region into the final state

 $${|\:\Psi\rangle}_f =  {  - {1\over {2\sqrt 2}} {|\: d\rangle|\:z\uparrow\rangle}} +$$

$${i\over {2\sqrt 2}} {|\:c\rangle|\: z\!\uparrow\rangle} + {i\over {\sqrt 2}} {|\:c\rangle|\: z\!\downarrow\rangle} \eqno(2)$$

The terms involving detection at C involve ambiguous states of the atom; we disregard these and focus our attention on the interesting case which is detection at D, represented by the second term in (2). When the photon component of this offer wave is absorbed by D, a photon CW is produced of the same initial amplitude, ${1\over 2} {\langle d|}$. The combined system OW also carries with it a factor of ${1\over \sqrt 2} $ given the original OW for the atom (which had an amplitude of ${1\over \sqrt 2}$ for that component of the state).\footnote{\normalsize In a private correspondence, A. Elitzur has raised the issue of a stationary superposed atom's OW as possibly needing some attention. It is possible that the OW and CW of massive particles are de Broglie phase waves. In the particle's rest frame, such waves propagate with infinite velocity. In other inertial frames, the phase waves propagate at speeds greater than $c$. Recall also that spin 1/2 particles described by the Dirac equation are undergoing ``zitterbewegung'' (zbw), a rapid oscillatory motion that is widely thought to be the source of spin. If, as in Schr\"{o}dinger's interpretation (1931), the oscillation is between positive and negative frequency solutions (OW and CW respectively), zbw may be telling us that spin 1/2 particles acquire their spin characteristics through a vacuum-mediated interweaving of OW and CW.}

The photon CW component propagates back along path $u$ through the two beam splitters, acquiring another factor of $1\over 2$ along the way, while the atom CW picks up another factor of ${1\over \sqrt 2} $ as it proceeds back to the source (which was prepared in state $|x\uparrow\rangle$),\footnote{\normalsize It is a basic postulate of TI that a CW, in order to interact with the source and thereby to be eligible for a transaction, must match the state of the OW emitted by the source. (See, e.g., Cramer (1986), p. 669, eq. 12). This reflects a natural symmetry with respect to the situation at the absorber, which emits a CW matching the state of the OW absorbed by it, and leads unambiguously to the perfect square symmetry expressed in the Born Rule. If we assume an idealized case in which an atom is emitted in the state $|x\uparrow\rangle$, then the returning CW, which is in an eigenstate of z-spin, will be attenuated by a factor of ${1\over \sqrt 2}$, as its $|x\downarrow\rangle$ component cannot interact with the source. Of course, the more realistic situation will be an atom emitted in some random spin state which will then need to be passed through a S-G device oriented along x, but the same principle applies: the source will only interact with the CW component which matches it.}
for a total CW amplitude at the photon and atom emitters of $[{1\over 4}] [{1\over 2}]  = { 1\over 8}$, in agreement with standard predictions. Note that
it is assumed that a measurement is made on the atom (or at least that it is ultimately absorbed) and an atomic CW is produced. The form of the combined OW ensures that a D transaction
can only occur in the presence of an atomic CW in the state $\langle z\!\uparrow, 0|$, which helps to explain why the atom's initial
superposition must be ``collapsed'' whenever the photon is detected at D.\vskip .5cm

{\bf 3. The Quantum Liar Experiment.}   \vskip .5cm

We now turn to a similar TI-based analysis of the QLE.
This experiment adds to the previous one a second atom prepared in the same way as the first, but with its intersecting box placed in the way of the other arm of the MZI, and corresponding to the state $\ketzd$ for the second atom. (Refer to Figure 2).\vskip 1cm

\unitlength 1mm 
\linethickness{0.4pt}
\ifx\plotpoint\undefined\newsavebox{\plotpoint}\fi 
\begin{picture}(129.5,107)(0,0)
\put(40.25,20.75){\framebox(53.25,53)[cc]{}}
\put(93.75,73.75){\line(1,0){27}}
\put(93.25,73.75){\line(0,1){24.75}}
\put(93.125,101.375){\oval(4.75,4.75)[t]}
\put(123.375,73.5){\oval(4.75,5)[r]}
\multiput(90.25,17.25)(.033653846,.037259615){208}{\line(0,1){.037259615}}
\multiput(36,70.25)(.038372093,.03372093){215}{\line(1,0){.038372093}}
\multiput(88.18,69.93)(.0411765,.0323529){17}{\line(1,0){.0411765}}
\multiput(89.58,71.03)(.0411765,.0323529){17}{\line(1,0){.0411765}}
\multiput(90.98,72.13)(.0411765,.0323529){17}{\line(1,0){.0411765}}
\multiput(92.38,73.23)(.0411765,.0323529){17}{\line(1,0){.0411765}}
\multiput(93.78,74.33)(.0411765,.0323529){17}{\line(1,0){.0411765}}
\multiput(95.18,75.43)(.0411765,.0323529){17}{\line(1,0){.0411765}}
\multiput(96.58,76.53)(.0411765,.0323529){17}{\line(1,0){.0411765}}
\multiput(97.98,77.63)(.0411765,.0323529){17}{\line(1,0){.0411765}}
\multiput(34.68,15.93)(.0378289,.0320724){19}{\line(1,0){.0378289}}
\multiput(36.117,17.148)(.0378289,.0320724){19}{\line(1,0){.0378289}}
\multiput(37.555,18.367)(.0378289,.0320724){19}{\line(1,0){.0378289}}
\multiput(38.992,19.586)(.0378289,.0320724){19}{\line(1,0){.0378289}}
\multiput(40.43,20.805)(.0378289,.0320724){19}{\line(1,0){.0378289}}
\multiput(41.867,22.023)(.0378289,.0320724){19}{\line(1,0){.0378289}}
\multiput(43.305,23.242)(.0378289,.0320724){19}{\line(1,0){.0378289}}
\multiput(44.742,24.461)(.0378289,.0320724){19}{\line(1,0){.0378289}}
\put(62.75,17){\dashbox{1}(6.75,7)[cc]{}}
\put(62.75,5.75){\dashbox{1}(6.75,7)[cc]{}}
\put(51.5,22.75){\makebox(0,0)[cc]{$v$}}
\put(44.25,33.75){\makebox(0,0)[cc]{$u$}}
\put(97.25,84.5){\makebox(0,0)[cc]{$d$}}
\put(104.75,70.75){\makebox(0,0)[cc]{$c$}}
\put(93,107){\makebox(0,0)[cc]{D}}
\put(129.5,73.75){\makebox(0,0)[cc]{C}}
\put(40.25,21){\line(-1,0){15}}
\put(22.25,21){\makebox(0,0)[cc]{L}}
\put(66.5,20.25){\makebox(0,0)[cc]{$\uparrow$}}
\put(66.5,8.75){\makebox(0,0)[cc]{$\downarrow$}}
\put(40.75,15.5){\makebox(0,0)[cc]{S1}}
\put(89.75,77.75){\makebox(0,0)[cc]{S2}}
\put(38.25,76.25){\makebox(0,0)[cc]{B}}
\put(96,19){\makebox(0,0)[cc]{A}}
\put(63.75,80.5){\dashbox{1}(6.75,7)[cc]{}}
\put(63.75,69.25){\dashbox{1}(6.75,7)[cc]{}}
\put(67.5,83.75){\makebox(0,0)[cc]{$\uparrow$}}
\put(67.5,72.25){\makebox(0,0)[cc]{$\downarrow$}}
\put(56.5,79){\makebox(0,0)[cc]{atom 2}}
\put(56,15.25){\makebox(0,0)[cc]{atom 1}}
\end{picture}

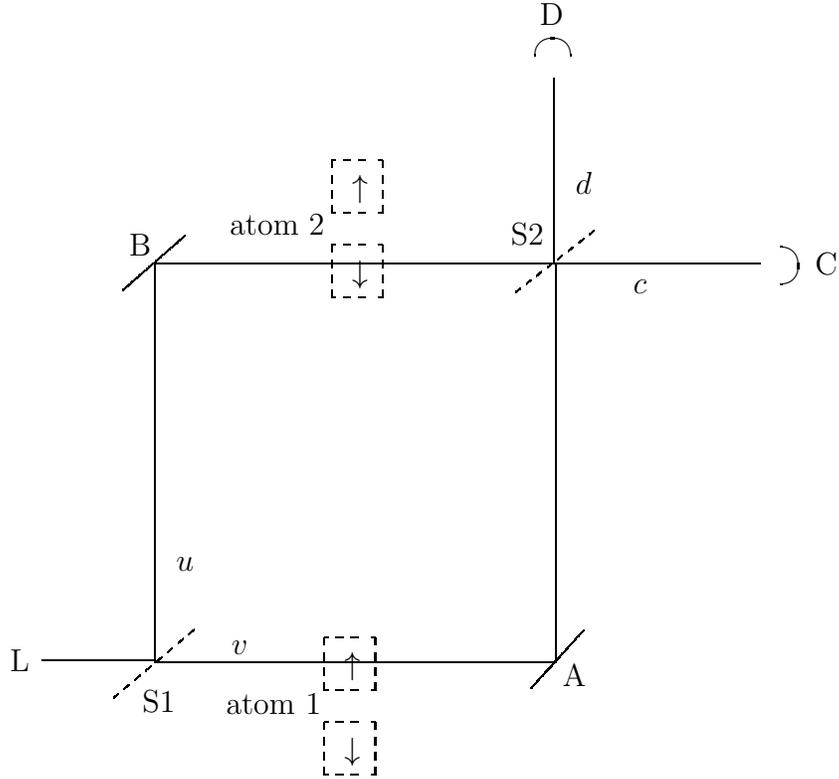
\normalsize \centerline{Figure 2. The Quantum Liar Experiment (QLE).}\vskip .5cm
\large

We first note that Elitzur and Dolev (2005) present this experiment as an example of what they describe as a situation ``analogous to a Schr\"{o}dinger cat found to be dead alongside scratches and droppings within the box that indicate that it has been alive all the time.'' [p. 341] In other words, according to the usual way of retrodicting trajectories for a photon based on the known state of the system and detections at one place or another, a seeming inconsistency arises within the description of the history of the system. However, we will see that under TI no such inconsistency arises.

The initial state of the system, in obvious notation, is\footnote{\normalsize{The factors of $i$ for the spin ``up'' states yield a more convenient form of the EPR state.}}:

$$|\Psi\rangle = |s\rangle \otimes {1\over \sqrt 2}  [\: i \ketzu_{\!\!1} + \ketzd_{\!\!1}\:] \otimes {1\over \sqrt 2} [\:i \ketzu_{\!\!2} + \ketzd_{\!\!2}\:] \eqno (3)$$

After the photon (or in this case OW) propagates through the first beam splitter S1, the state is:

$$|\Psi\rangle = {1\over {2\sqrt 2}}[ i |u\rangle + |v\rangle ] \otimes   [\: i \ketzu_{\!\!1} + \ketzd_{\!\!1}\:] \otimes  [\:i \ketzu_{\!\!2} + \ketzd_{\!\!2}\:] \eqno (4)$$

This product yields eight terms corresponding to eight total system OW components. In the following we adopt the shorthand $\ketzu_{\!\!1}\ketzu_{\!\!2} = \bsu$, etc. The photon will be absorbed in transactions corresponding to OW components with atomic states $\ud$ and for single components $\ketu \bsd$ and $\ketv \bsu$ in a process very similar to the single-atom case above.

For now, we consider only the remainder, reviewing the conventional account  :

$$|\Psi\rangle ={1\over {2\sqrt 2}} [-i \ketu \bsu - \ketu\du + i \ketv \du
+ \ketv\bsd ] \eqno(5)$$

After the photon passes through the second beam splitter, with the evolution
$$\ketv \rightarrow {1\over\sqrt 2} {(\ketd + i\ketc)}, \hskip .5cm
\ketu \rightarrow {1\over\sqrt 2} {(\ketc + i\ketd)}, $$

the state becomes

$$|\Psi\rangle = \frac{1}{4} (\ketd \bsu + \ketd \bsd + i\ketc \bsd -i\ketc\bsu - 2\ketc\du). \eqno (5)$$

If we now select only those photons detected at D, indicating the presence of an atom in an intersecting box, we have:

$$|\Psi\rangle_D = \frac{1}{4} \ketd (\bsu + \bsd) \eqno (6)$$

That is, the atoms are now entangled in an EPR state, where, in the words of EDZ, ``the only common event in [the atoms'] past is the single photon that has `visited' both of them.'' (2002, 454) EDZ then go on to present a version of the experiment in which even the common past event seems to be eliminated, by exploiting the Hanbury-Twiss effect (to be discussed in more detail in part 4).

There are now some apparently strange and paradoxical implications of this result for the atoms for which photons were detected at D. If we choose to open one of the boxes, this will constitute a measurement of the $z$ spin of that atom, and according to $|\Psi\rangle_D$, the other atom must always be found with the same spin. The usual account is that this means that one atom blocks one of the MZI's arms while the other lies outside of it, and therefore  the photon must definitely have traveled along the unblocked arm; but that would mean that it didn't interact with the atom on the other arm, so how could it have brought about any kind of correlation between the atoms?

Elitzur and Dolev refer to this as the ``quantum liar'' experiment because, in their words: ``The very fact that one atom is positioned in a place that seems to preclude its interaction with the other atom leads to its being affected by that other atom. This is logically equivalent to the statement: `This sentence has never been written.''' (ED p. 344)

Moreover, if we choose to measure the atoms' spins along other directions (by bringing the boxes back together and then separating them along other spin directions), we find that such outcomes violate the Bell inequality. Therefore, neither atom was really in a definite $z$ spin state, so there was no definite ``silent detector'' (blocking box) and the photon could not have gone only along one path or another. This contradictory situation is analogous to the schizophrenic Schr\"{o}dinger cat who, upon opening the box, is found to be dead but with evidence
of a cat having been alive for the entire time (scratches and droppings in the box).

Let us now examine the QLE under TI. We will follow the notational convention in Cramer (2006), which traces an OW component from the source to a detector, and a CW from a detector back to the source. (The photon source, a laser, is denoted by L. Reflections at beam splitters resulting in a phase change of $\frac{\pi}{2}$ are underlined.)

First we limit the discussion to OW resulting in photon detection at D. Such detections arise from total system OW components with matching spin values, since the photon OW corresponding to the remaining nonmatching atomic spin state ($\du$) cancel out at D due to destructive interference:

$$ |L-\underline{S1}-A-\underline{S2}-D\rangle + |L-S1-B-S2-D\rangle = 0 \eqno(7)$$

So the only offer wave components that can give rise to transactions are

$$ |L-\underline{S1}-B-\underline{S2}-D\rangle \bsu\eqno (8a)$$

and

$$ |L-S1-A-S2-D\rangle \bsd \eqno(8b)$$

Now, under TI we have to consider the atomic CW that are present in cases where the photon is detected at D (see Figure 3). First, suppose that the box along path $v$ is opened after a photon is detected at D, thus measuring the $z$ spin of atom 1 (and that atom 2 is also detected/absorbed at some point). A detection of atom 1 revealing its $z$ spin generates CWs of the form $\brazu$, $\brazd$. The available total system offer waves constrain possible transactions to those in which CWs from both atoms match, with the photon OW corresponding to the states of the atoms as in (8a,b).
\vskip .5cm
\unitlength 1mm 
\linethickness{0.4pt}
\ifx\plotpoint\undefined\newsavebox{\plotpoint}\fi 
\begin{picture}(86.75,105.75)(0,0)
\multiput(34.75,44)(.0337209302,-.0391472868){645}{\line(0,-1){.0391472868}}
\multiput(35,44.25)(.03370418848,.04155759162){764}{\line(0,1){.04155759162}}
\put(24,96){\dashbox{1}(17.25,5.5)[cc]{}}
\put(65.25,96){\dashbox{1}(17.25,5.5)[cc]{}}
\put(79.75,79.75){\vector(0,-1){.07}}\put(79.43,96.18){\line(0,-1){.9706}}
\put(79.459,94.239){\line(0,-1){.9706}}
\put(79.489,92.297){\line(0,-1){.9706}}
\put(79.518,90.356){\line(0,-1){.9706}}
\put(79.547,88.415){\line(0,-1){.9706}}
\put(79.577,86.474){\line(0,-1){.9706}}
\put(79.606,84.533){\line(0,-1){.9706}}
\put(79.636,82.591){\line(0,-1){.9706}}
\put(79.665,80.65){\line(0,-1){.9706}}
\put(79.694,78.709){\line(0,-1){.9706}}
\put(79.724,76.768){\line(0,-1){.9706}}
\put(79.753,74.827){\line(0,-1){.9706}}
\put(79.783,72.886){\line(0,-1){.9706}}
\put(79.812,70.944){\line(0,-1){.9706}}
\put(79.841,69.003){\line(0,-1){.9706}}
\put(79.871,67.062){\line(0,-1){.9706}}
\put(79.9,65.121){\line(0,-1){.9706}}
\put(39,80){\vector(0,-1){.07}}\put(38.68,96.43){\line(0,-1){.9706}}
\put(38.709,94.489){\line(0,-1){.9706}}
\put(38.739,92.547){\line(0,-1){.9706}}
\put(38.768,90.606){\line(0,-1){.9706}}
\put(38.797,88.665){\line(0,-1){.9706}}
\put(38.827,86.724){\line(0,-1){.9706}}
\put(38.856,84.783){\line(0,-1){.9706}}
\put(38.886,82.841){\line(0,-1){.9706}}
\put(38.915,80.9){\line(0,-1){.9706}}
\put(38.944,78.959){\line(0,-1){.9706}}
\put(38.974,77.018){\line(0,-1){.9706}}
\put(39.003,75.077){\line(0,-1){.9706}}
\put(39.033,73.136){\line(0,-1){.9706}}
\put(39.062,71.194){\line(0,-1){.9706}}
\put(39.091,69.253){\line(0,-1){.9706}}
\put(39.121,67.312){\line(0,-1){.9706}}
\put(39.15,65.371){\line(0,-1){.9706}}
\multiput(76.18,43.68)(-.0332447,.0337766){20}{\line(0,1){.0337766}}
\multiput(74.85,45.031)(-.0332447,.0337766){20}{\line(0,1){.0337766}}
\multiput(73.52,46.382)(-.0332447,.0337766){20}{\line(0,1){.0337766}}
\multiput(72.19,47.733)(-.0332447,.0337766){20}{\line(0,1){.0337766}}
\multiput(70.861,49.084)(-.0332447,.0337766){20}{\line(0,1){.0337766}}
\multiput(69.531,50.435)(-.0332447,.0337766){20}{\line(0,1){.0337766}}
\multiput(68.201,51.786)(-.0332447,.0337766){20}{\line(0,1){.0337766}}
\multiput(66.871,53.137)(-.0332447,.0337766){20}{\line(0,1){.0337766}}
\multiput(65.541,54.488)(-.0332447,.0337766){20}{\line(0,1){.0337766}}
\multiput(64.212,55.839)(-.0332447,.0337766){20}{\line(0,1){.0337766}}
\multiput(62.882,57.19)(-.0332447,.0337766){20}{\line(0,1){.0337766}}
\multiput(61.552,58.541)(-.0332447,.0337766){20}{\line(0,1){.0337766}}
\multiput(60.222,59.892)(-.0332447,.0337766){20}{\line(0,1){.0337766}}
\multiput(58.892,61.244)(-.0332447,.0337766){20}{\line(0,1){.0337766}}
\multiput(57.563,62.595)(-.0332447,.0337766){20}{\line(0,1){.0337766}}
\multiput(56.233,63.946)(-.0332447,.0337766){20}{\line(0,1){.0337766}}
\multiput(54.903,65.297)(-.0332447,.0337766){20}{\line(0,1){.0337766}}
\multiput(53.573,66.648)(-.0332447,.0337766){20}{\line(0,1){.0337766}}
\multiput(52.244,67.999)(-.0332447,.0337766){20}{\line(0,1){.0337766}}
\multiput(50.914,69.35)(-.0332447,.0337766){20}{\line(0,1){.0337766}}
\multiput(49.584,70.701)(-.0332447,.0337766){20}{\line(0,1){.0337766}}
\multiput(48.254,72.052)(-.0332447,.0337766){20}{\line(0,1){.0337766}}
\multiput(46.924,73.403)(-.0332447,.0337766){20}{\line(0,1){.0337766}}
\multiput(45.595,74.754)(-.0332447,.0337766){20}{\line(0,1){.0337766}}
\multiput(55.93,19.18)(.0326797,.0408497){18}{\line(0,1){.0408497}}
\multiput(57.106,20.65)(.0326797,.0408497){18}{\line(0,1){.0408497}}
\multiput(58.283,22.121)(.0326797,.0408497){18}{\line(0,1){.0408497}}
\multiput(59.459,23.591)(.0326797,.0408497){18}{\line(0,1){.0408497}}
\multiput(60.636,25.062)(.0326797,.0408497){18}{\line(0,1){.0408497}}
\multiput(61.812,26.533)(.0326797,.0408497){18}{\line(0,1){.0408497}}
\multiput(62.989,28.003)(.0326797,.0408497){18}{\line(0,1){.0408497}}
\multiput(64.165,29.474)(.0326797,.0408497){18}{\line(0,1){.0408497}}
\multiput(65.341,30.944)(.0326797,.0408497){18}{\line(0,1){.0408497}}
\multiput(66.518,32.415)(.0326797,.0408497){18}{\line(0,1){.0408497}}
\multiput(67.694,33.886)(.0326797,.0408497){18}{\line(0,1){.0408497}}
\multiput(68.871,35.356)(.0326797,.0408497){18}{\line(0,1){.0408497}}
\multiput(70.047,36.827)(.0326797,.0408497){18}{\line(0,1){.0408497}}
\multiput(71.224,38.297)(.0326797,.0408497){18}{\line(0,1){.0408497}}
\multiput(72.4,39.768)(.0326797,.0408497){18}{\line(0,1){.0408497}}
\multiput(73.577,41.239)(.0326797,.0408497){18}{\line(0,1){.0408497}}
\multiput(74.753,42.709)(.0326797,.0408497){18}{\line(0,1){.0408497}}
\multiput(56,19.25)(-.03358209,-.039179104){134}{\line(0,-1){.039179104}}
\put(80,63.25){\vector(0,1){.07}}\put(81.18,29.43){\line(0,1){.9926}}
\put(81.106,31.415){\line(0,1){.9926}}
\put(81.033,33.4){\line(0,1){.9926}}
\put(80.959,35.386){\line(0,1){.9926}}
\put(80.886,37.371){\line(0,1){.9926}}
\put(80.812,39.356){\line(0,1){.9926}}
\put(80.739,41.341){\line(0,1){.9926}}
\put(80.665,43.327){\line(0,1){.9926}}
\put(80.591,45.312){\line(0,1){.9926}}
\put(80.518,47.297){\line(0,1){.9926}}
\put(80.444,49.283){\line(0,1){.9926}}
\put(80.371,51.268){\line(0,1){.9926}}
\put(80.297,53.253){\line(0,1){.9926}}
\put(80.224,55.239){\line(0,1){.9926}}
\put(80.15,57.224){\line(0,1){.9926}}
\put(80.077,59.209){\line(0,1){.9926}}
\put(80.003,61.194){\line(0,1){.9926}}
\put(39.25,64.75){\vector(0,1){.07}}\put(40.18,30.18){\line(0,1){.9857}}
\put(40.123,32.151){\line(0,1){.9857}}
\put(40.065,34.123){\line(0,1){.9857}}
\put(40.008,36.094){\line(0,1){.9857}}
\put(39.951,38.065){\line(0,1){.9857}}
\put(39.894,40.037){\line(0,1){.9857}}
\put(39.837,42.008){\line(0,1){.9857}}
\put(39.78,43.98){\line(0,1){.9857}}
\put(39.723,45.951){\line(0,1){.9857}}
\put(39.665,47.923){\line(0,1){.9857}}
\put(39.608,49.894){\line(0,1){.9857}}
\put(39.551,51.865){\line(0,1){.9857}}
\put(39.494,53.837){\line(0,1){.9857}}
\put(39.437,55.808){\line(0,1){.9857}}
\put(39.38,57.78){\line(0,1){.9857}}
\put(39.323,59.751){\line(0,1){.9857}}
\put(39.265,61.723){\line(0,1){.9857}}
\put(39.208,63.694){\line(0,1){.9857}}
\put(73.75,105.5){\makebox(0,0)[cc]{Z}}
\put(34.5,105.75){\makebox(0,0)[cc]{Z}}
\put(81.5,30){\vector(1,3){.07}}\multiput(75.43,10.93)(.031746,.100529){9}{\line(0,1){.100529}}
\multiput(76.001,12.739)(.031746,.100529){9}{\line(0,1){.100529}}
\multiput(76.573,14.549)(.031746,.100529){9}{\line(0,1){.100529}}
\multiput(77.144,16.358)(.031746,.100529){9}{\line(0,1){.100529}}
\multiput(77.715,18.168)(.031746,.100529){9}{\line(0,1){.100529}}
\multiput(78.287,19.977)(.031746,.100529){9}{\line(0,1){.100529}}
\multiput(78.858,21.787)(.031746,.100529){9}{\line(0,1){.100529}}
\multiput(79.43,23.596)(.031746,.100529){9}{\line(0,1){.100529}}
\multiput(80.001,25.406)(.031746,.100529){9}{\line(0,1){.100529}}
\multiput(80.573,27.215)(.031746,.100529){9}{\line(0,1){.100529}}
\multiput(81.144,29.025)(.031746,.100529){9}{\line(0,1){.100529}}
\put(69.75,29.75){\vector(-1,4){.07}}\multiput(75,11)(-.033653846,.120192308){156}{\line(0,1){.120192308}}
\put(28.25,29.75){\vector(-1,4){.07}}\multiput(33.5,11)(-.033653846,.120192308){156}{\line(0,1){.120192308}}
\put(75.25,5.25){\vector(0,1){5.75}}
\put(33.75,5.25){\vector(0,1){5.75}}
\put(69.75,29.75){\vector(0,1){31.25}}
\put(28.25,29.75){\vector(0,1){31.25}}
\put(69.5,78.375){\vector(0,-1){.07}}\put(69.5,95.75){\line(0,-1){34.75}}
\put(28,78.375){\vector(0,-1){.07}}\put(28,95.75){\line(0,-1){34.75}}
\put(40.25,30.5){\vector(1,3){.07}}\multiput(33.93,10.93)(.033069,.103175){9}{\line(0,1){.103175}}
\multiput(34.525,12.787)(.033069,.103175){9}{\line(0,1){.103175}}
\multiput(35.12,14.644)(.033069,.103175){9}{\line(0,1){.103175}}
\multiput(35.715,16.501)(.033069,.103175){9}{\line(0,1){.103175}}
\multiput(36.311,18.358)(.033069,.103175){9}{\line(0,1){.103175}}
\multiput(36.906,20.215)(.033069,.103175){9}{\line(0,1){.103175}}
\multiput(37.501,22.073)(.033069,.103175){9}{\line(0,1){.103175}}
\multiput(38.096,23.93)(.033069,.103175){9}{\line(0,1){.103175}}
\multiput(38.692,25.787)(.033069,.103175){9}{\line(0,1){.103175}}
\multiput(39.287,27.644)(.033069,.103175){9}{\line(0,1){.103175}}
\multiput(39.882,29.501)(.033069,.103175){9}{\line(0,1){.103175}}
\put(44.5,78.5){\makebox(0,0)[cc]{D}}
\put(61.5,78.75){\makebox(0,0)[cc]{C}}
\put(50.5,13.25){\makebox(0,0)[cc]{L}}
\put(68.75,98.75){\makebox(0,0)[cc]{+}}
\put(27.75,98.5){\makebox(0,0)[cc]{+}}
\put(75.25,1.5){\makebox(0,0)[cc]{atom 1}}
\put(33.75,2){\makebox(0,0)[cc]{atom 2}}
\thicklines
\put(78.5,32.5){\framebox(5.75,5.75)[cc]{}}
\put(67,32.5){\framebox(5.75,5.75)[cc]{}}
\put(37.25,33){\framebox(5.75,5.75)[cc]{}}
\put(25.25,32.75){\framebox(5.75,5.75)[cc]{}}
\put(22.75,35.75){\makebox(0,0)[cc]{+}}
\put(38,98.75){\makebox(0,0)[cc]{-}}
\put(79.25,99){\makebox(0,0)[cc]{-}}
\put(45.25,36.5){\makebox(0,0)[cc]{-}}
\put(86.75,36.5){\makebox(0,0)[cc]{-}}
\put(64.75,35.75){\makebox(0,0)[cc]{+}}
\put(48,24.75){\makebox(0,0)[cc]{$u$}}
\put(64,24.5){\makebox(0,0)[cc]{$v$}}
\end{picture}

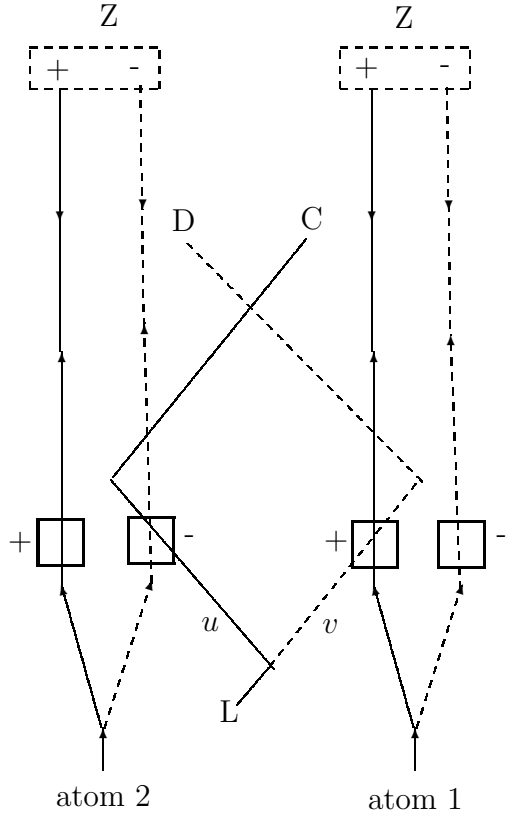
\normalsize\centerline{Figure 3. The QLE with a measurement of the atoms' spins along $z$.}
 \vskip .5cm
 \large

Thus, if it turns out that atom 1 was in the intersecting box, the realized transaction is the one with an atomic component corresponding to $\bsu$, and the photon component corresponding to path $u$ (8a). The situation is exactly reversed if atom 1 is found not to be in the box when it is opened; in this case the realized transaction is the one corresponding to path $v$ and atomic state $\bsd$ (8b).

On the other hand, what happens if we decide not to open a box, but instead bring them back together and measure the atomic spins along some other direction, say $y$ (using the usual Stern-Gerlach apparatus)? (See Figure 4). This will generate atomic CW of the form $\brayu$, $\brayd$. Each of these contains equal amounts of the $z$-spin eigenstates, which will separate according to the earlier $z$-oriented Stern-Gerlach field at the location of the MZI.  For the atoms to be found in either eigenstate of spin along $y$, equal OW and CW components of each atom's spin along $z$ is required, so possible transactions involve superpositions of the states (8). These will yield no ``fact of the matter'' either for the atoms' whereabouts in the boxes nor for the photon's whereabouts on either path.\vskip .5cm

\unitlength 1mm 
\linethickness{0.4pt}
\ifx\plotpoint\undefined\newsavebox{\plotpoint}\fi 
\begin{picture}(86.75,104.75)(0,0)
\multiput(34.75,44)(.0337209302,-.0391472868){645}{\line(0,-1){.0391472868}}
\multiput(35,44.25)(.03370418848,.04155759162){764}{\line(0,1){.04155759162}}
\put(24,96){\dashbox{1}(17.25,5.5)[cc]{}}
\put(65.25,96){\dashbox{1}(17.25,5.5)[cc]{}}
\multiput(76.18,43.68)(-.0332447,.0337766){20}{\line(0,1){.0337766}}
\multiput(74.85,45.031)(-.0332447,.0337766){20}{\line(0,1){.0337766}}
\multiput(73.52,46.382)(-.0332447,.0337766){20}{\line(0,1){.0337766}}
\multiput(72.19,47.733)(-.0332447,.0337766){20}{\line(0,1){.0337766}}
\multiput(70.861,49.084)(-.0332447,.0337766){20}{\line(0,1){.0337766}}
\multiput(69.531,50.435)(-.0332447,.0337766){20}{\line(0,1){.0337766}}
\multiput(68.201,51.786)(-.0332447,.0337766){20}{\line(0,1){.0337766}}
\multiput(66.871,53.137)(-.0332447,.0337766){20}{\line(0,1){.0337766}}
\multiput(65.541,54.488)(-.0332447,.0337766){20}{\line(0,1){.0337766}}
\multiput(64.212,55.839)(-.0332447,.0337766){20}{\line(0,1){.0337766}}
\multiput(62.882,57.19)(-.0332447,.0337766){20}{\line(0,1){.0337766}}
\multiput(61.552,58.541)(-.0332447,.0337766){20}{\line(0,1){.0337766}}
\multiput(60.222,59.892)(-.0332447,.0337766){20}{\line(0,1){.0337766}}
\multiput(58.892,61.244)(-.0332447,.0337766){20}{\line(0,1){.0337766}}
\multiput(57.563,62.595)(-.0332447,.0337766){20}{\line(0,1){.0337766}}
\multiput(56.233,63.946)(-.0332447,.0337766){20}{\line(0,1){.0337766}}
\multiput(54.903,65.297)(-.0332447,.0337766){20}{\line(0,1){.0337766}}
\multiput(53.573,66.648)(-.0332447,.0337766){20}{\line(0,1){.0337766}}
\multiput(52.244,67.999)(-.0332447,.0337766){20}{\line(0,1){.0337766}}
\multiput(50.914,69.35)(-.0332447,.0337766){20}{\line(0,1){.0337766}}
\multiput(49.584,70.701)(-.0332447,.0337766){20}{\line(0,1){.0337766}}
\multiput(48.254,72.052)(-.0332447,.0337766){20}{\line(0,1){.0337766}}
\multiput(46.924,73.403)(-.0332447,.0337766){20}{\line(0,1){.0337766}}
\multiput(45.595,74.754)(-.0332447,.0337766){20}{\line(0,1){.0337766}}
\multiput(55.93,19.18)(.0326797,.0408497){18}{\line(0,1){.0408497}}
\multiput(57.106,20.65)(.0326797,.0408497){18}{\line(0,1){.0408497}}
\multiput(58.283,22.121)(.0326797,.0408497){18}{\line(0,1){.0408497}}
\multiput(59.459,23.591)(.0326797,.0408497){18}{\line(0,1){.0408497}}
\multiput(60.636,25.062)(.0326797,.0408497){18}{\line(0,1){.0408497}}
\multiput(61.812,26.533)(.0326797,.0408497){18}{\line(0,1){.0408497}}
\multiput(62.989,28.003)(.0326797,.0408497){18}{\line(0,1){.0408497}}
\multiput(64.165,29.474)(.0326797,.0408497){18}{\line(0,1){.0408497}}
\multiput(65.341,30.944)(.0326797,.0408497){18}{\line(0,1){.0408497}}
\multiput(66.518,32.415)(.0326797,.0408497){18}{\line(0,1){.0408497}}
\multiput(67.694,33.886)(.0326797,.0408497){18}{\line(0,1){.0408497}}
\multiput(68.871,35.356)(.0326797,.0408497){18}{\line(0,1){.0408497}}
\multiput(70.047,36.827)(.0326797,.0408497){18}{\line(0,1){.0408497}}
\multiput(71.224,38.297)(.0326797,.0408497){18}{\line(0,1){.0408497}}
\multiput(72.4,39.768)(.0326797,.0408497){18}{\line(0,1){.0408497}}
\multiput(73.577,41.239)(.0326797,.0408497){18}{\line(0,1){.0408497}}
\multiput(74.753,42.709)(.0326797,.0408497){18}{\line(0,1){.0408497}}
\multiput(56,19.25)(-.03358209,-.039179104){134}{\line(0,-1){.039179104}}
\put(81.5,30){\vector(1,3){.07}}\multiput(75.43,10.93)(.031746,.100529){9}{\line(0,1){.100529}}
\multiput(76.001,12.739)(.031746,.100529){9}{\line(0,1){.100529}}
\multiput(76.573,14.549)(.031746,.100529){9}{\line(0,1){.100529}}
\multiput(77.144,16.358)(.031746,.100529){9}{\line(0,1){.100529}}
\multiput(77.715,18.168)(.031746,.100529){9}{\line(0,1){.100529}}
\multiput(78.287,19.977)(.031746,.100529){9}{\line(0,1){.100529}}
\multiput(78.858,21.787)(.031746,.100529){9}{\line(0,1){.100529}}
\multiput(79.43,23.596)(.031746,.100529){9}{\line(0,1){.100529}}
\multiput(80.001,25.406)(.031746,.100529){9}{\line(0,1){.100529}}
\multiput(80.573,27.215)(.031746,.100529){9}{\line(0,1){.100529}}
\multiput(81.144,29.025)(.031746,.100529){9}{\line(0,1){.100529}}
\put(69.75,29.75){\vector(-1,4){.07}}\multiput(75,11)(-.033653846,.120192308){156}{\line(0,1){.120192308}}
\put(28.25,29.75){\vector(-1,4){.07}}\multiput(33.5,11)(-.033653846,.120192308){156}{\line(0,1){.120192308}}
\put(40.25,30.5){\vector(1,3){.07}}\multiput(33.93,10.93)(.033069,.103175){9}{\line(0,1){.103175}}
\multiput(34.525,12.787)(.033069,.103175){9}{\line(0,1){.103175}}
\multiput(35.12,14.644)(.033069,.103175){9}{\line(0,1){.103175}}
\multiput(35.715,16.501)(.033069,.103175){9}{\line(0,1){.103175}}
\multiput(36.311,18.358)(.033069,.103175){9}{\line(0,1){.103175}}
\multiput(36.906,20.215)(.033069,.103175){9}{\line(0,1){.103175}}
\multiput(37.501,22.073)(.033069,.103175){9}{\line(0,1){.103175}}
\multiput(38.096,23.93)(.033069,.103175){9}{\line(0,1){.103175}}
\multiput(38.692,25.787)(.033069,.103175){9}{\line(0,1){.103175}}
\multiput(39.287,27.644)(.033069,.103175){9}{\line(0,1){.103175}}
\multiput(39.882,29.501)(.033069,.103175){9}{\line(0,1){.103175}}
\put(44.5,78.5){\makebox(0,0)[cc]{D}}
\put(61.5,78.75){\makebox(0,0)[cc]{C}}
\put(50.5,13.25){\makebox(0,0)[cc]{L}}
\put(68.75,98.75){\makebox(0,0)[cc]{+}}
\put(27.75,98.5){\makebox(0,0)[cc]{+}}
\put(75.25,1.5){\makebox(0,0)[cc]{atom 1}}
\put(33.75,2){\makebox(0,0)[cc]{atom 2}}
\thicklines
\put(78.5,32.5){\framebox(5.75,5.75)[cc]{}}
\put(67,32.5){\framebox(5.75,5.75)[cc]{}}
\put(37.25,33){\framebox(5.75,5.75)[cc]{}}
\put(25.25,32.75){\framebox(5.75,5.75)[cc]{}}
\put(22.75,35.75){\makebox(0,0)[cc]{+}}
\put(38,98.75){\makebox(0,0)[cc]{-}}
\put(79.25,99){\makebox(0,0)[cc]{-}}
\put(45.25,36.5){\makebox(0,0)[cc]{-}}
\put(86.75,36.5){\makebox(0,0)[cc]{-}}
\put(64.75,35.75){\makebox(0,0)[cc]{+}}
\thinlines
\put(73.75,50){\vector(-2,3){.07}}\multiput(81.43,38.93)(-.0322917,.0458333){16}{\line(0,1){.0458333}}
\multiput(80.396,40.396)(-.0322917,.0458333){16}{\line(0,1){.0458333}}
\multiput(79.363,41.863)(-.0322917,.0458333){16}{\line(0,1){.0458333}}
\multiput(78.33,43.33)(-.0322917,.0458333){16}{\line(0,1){.0458333}}
\multiput(77.296,44.796)(-.0322917,.0458333){16}{\line(0,1){.0458333}}
\multiput(76.263,46.263)(-.0322917,.0458333){16}{\line(0,1){.0458333}}
\multiput(75.23,47.73)(-.0322917,.0458333){16}{\line(0,1){.0458333}}
\multiput(74.196,49.196)(-.0322917,.0458333){16}{\line(0,1){.0458333}}
\put(73.75,48.75){\vector(1,2){.07}}\multiput(68.25,39)(.033536585,.05945122){164}{\line(0,1){.05945122}}
\put(78.25,80.75){\vector(0,-1){.07}}\multiput(77.93,93.68)(.01786,-.92857){15}{{\rule{.4pt}{.4pt}}}
\put(67.75,81){\vector(0,-1){.07}}\multiput(67.68,94.18)(0,-.94643){15}{{\rule{.4pt}{.4pt}}}
\put(81.5,43.5){\vector(0,-1){.07}}\put(78.93,79.68){\line(0,-1){.9797}}
\put(79.065,77.72){\line(0,-1){.9797}}
\put(79.2,75.761){\line(0,-1){.9797}}
\put(79.335,73.801){\line(0,-1){.9797}}
\put(79.47,71.842){\line(0,-1){.9797}}
\put(79.605,69.882){\line(0,-1){.9797}}
\put(79.741,67.923){\line(0,-1){.9797}}
\put(79.876,65.963){\line(0,-1){.9797}}
\put(80.011,64.004){\line(0,-1){.9797}}
\put(80.146,62.045){\line(0,-1){.9797}}
\put(80.281,60.085){\line(0,-1){.9797}}
\put(80.416,58.126){\line(0,-1){.9797}}
\put(80.551,56.166){\line(0,-1){.9797}}
\put(80.686,54.207){\line(0,-1){.9797}}
\put(80.822,52.247){\line(0,-1){.9797}}
\put(80.957,50.288){\line(0,-1){.9797}}
\put(81.092,48.328){\line(0,-1){.9797}}
\put(81.227,46.369){\line(0,-1){.9797}}
\put(81.362,44.409){\line(0,-1){.9797}}
\put(80.5,42.75){\vector(1,-3){.07}}\multiput(67.93,79.68)(.03125,-.0925){10}{\line(0,-1){.0925}}
\multiput(68.555,77.83)(.03125,-.0925){10}{\line(0,-1){.0925}}
\multiput(69.18,75.98)(.03125,-.0925){10}{\line(0,-1){.0925}}
\multiput(69.805,74.13)(.03125,-.0925){10}{\line(0,-1){.0925}}
\multiput(70.43,72.28)(.03125,-.0925){10}{\line(0,-1){.0925}}
\multiput(71.055,70.43)(.03125,-.0925){10}{\line(0,-1){.0925}}
\multiput(71.68,68.58)(.03125,-.0925){10}{\line(0,-1){.0925}}
\multiput(72.305,66.73)(.03125,-.0925){10}{\line(0,-1){.0925}}
\multiput(72.93,64.88)(.03125,-.0925){10}{\line(0,-1){.0925}}
\multiput(73.555,63.03)(.03125,-.0925){10}{\line(0,-1){.0925}}
\multiput(74.18,61.18)(.03125,-.0925){10}{\line(0,-1){.0925}}
\multiput(74.805,59.33)(.03125,-.0925){10}{\line(0,-1){.0925}}
\multiput(75.43,57.48)(.03125,-.0925){10}{\line(0,-1){.0925}}
\multiput(76.055,55.63)(.03125,-.0925){10}{\line(0,-1){.0925}}
\multiput(76.68,53.78)(.03125,-.0925){10}{\line(0,-1){.0925}}
\multiput(77.305,51.93)(.03125,-.0925){10}{\line(0,-1){.0925}}
\multiput(77.93,50.08)(.03125,-.0925){10}{\line(0,-1){.0925}}
\multiput(78.555,48.23)(.03125,-.0925){10}{\line(0,-1){.0925}}
\multiput(79.18,46.38)(.03125,-.0925){10}{\line(0,-1){.0925}}
\multiput(79.805,44.53)(.03125,-.0925){10}{\line(0,-1){.0925}}
\put(78.25,81){\vector(-1,-4){10.25}}
\put(67.75,79){\vector(0,-1){36}}
\thicklines
\put(74,86.5){\vector(0,1){.07}}\multiput(74.25,49.5)(-.03125,4.625){8}{\line(0,1){4.625}}
\put(75.5,5.25){\vector(0,1){5.25}}
\put(77.5,94.75){\vector(1,3){.07}}\multiput(74.18,86.43)(.325,.825){11}{{\rule{.8pt}{.8pt}}}
\put(69,94.75){\vector(-2,3){.07}}\multiput(74.18,85.93)(-.47727,.79545){12}{{\rule{.8pt}{.8pt}}}
\thinlines
\put(33,50.25){\vector(-2,3){.07}}\multiput(40.68,39.18)(-.0322917,.0458333){16}{\line(0,1){.0458333}}
\multiput(39.646,40.646)(-.0322917,.0458333){16}{\line(0,1){.0458333}}
\multiput(38.613,42.113)(-.0322917,.0458333){16}{\line(0,1){.0458333}}
\multiput(37.58,43.58)(-.0322917,.0458333){16}{\line(0,1){.0458333}}
\multiput(36.546,45.046)(-.0322917,.0458333){16}{\line(0,1){.0458333}}
\multiput(35.513,46.513)(-.0322917,.0458333){16}{\line(0,1){.0458333}}
\multiput(34.48,47.98)(-.0322917,.0458333){16}{\line(0,1){.0458333}}
\multiput(33.446,49.446)(-.0322917,.0458333){16}{\line(0,1){.0458333}}
\put(33,49){\vector(1,2){.07}}\multiput(27.5,39.25)(.033536585,.05945122){164}{\line(0,1){.05945122}}
\put(37.5,81.75){\vector(0,-1){.07}}\multiput(37.18,94.68)(.01786,-.92857){15}{{\rule{.4pt}{.4pt}}}
\put(27,82){\vector(0,-1){.07}}\multiput(26.93,95.18)(0,-.94643){15}{{\rule{.4pt}{.4pt}}}
\put(40.75,44.5){\vector(0,-1){.07}}\put(38.18,80.68){\line(0,-1){.9797}}
\put(38.315,78.72){\line(0,-1){.9797}}
\put(38.45,76.761){\line(0,-1){.9797}}
\put(38.585,74.801){\line(0,-1){.9797}}
\put(38.72,72.842){\line(0,-1){.9797}}
\put(38.855,70.882){\line(0,-1){.9797}}
\put(38.991,68.923){\line(0,-1){.9797}}
\put(39.126,66.963){\line(0,-1){.9797}}
\put(39.261,65.004){\line(0,-1){.9797}}
\put(39.396,63.045){\line(0,-1){.9797}}
\put(39.531,61.085){\line(0,-1){.9797}}
\put(39.666,59.126){\line(0,-1){.9797}}
\put(39.801,57.166){\line(0,-1){.9797}}
\put(39.936,55.207){\line(0,-1){.9797}}
\put(40.072,53.247){\line(0,-1){.9797}}
\put(40.207,51.288){\line(0,-1){.9797}}
\put(40.342,49.328){\line(0,-1){.9797}}
\put(40.477,47.369){\line(0,-1){.9797}}
\put(40.612,45.409){\line(0,-1){.9797}}
\put(39.75,43.75){\vector(1,-3){.07}}\multiput(27.18,80.68)(.03125,-.0925){10}{\line(0,-1){.0925}}
\multiput(27.805,78.83)(.03125,-.0925){10}{\line(0,-1){.0925}}
\multiput(28.43,76.98)(.03125,-.0925){10}{\line(0,-1){.0925}}
\multiput(29.055,75.13)(.03125,-.0925){10}{\line(0,-1){.0925}}
\multiput(29.68,73.28)(.03125,-.0925){10}{\line(0,-1){.0925}}
\multiput(30.305,71.43)(.03125,-.0925){10}{\line(0,-1){.0925}}
\multiput(30.93,69.58)(.03125,-.0925){10}{\line(0,-1){.0925}}
\multiput(31.555,67.73)(.03125,-.0925){10}{\line(0,-1){.0925}}
\multiput(32.18,65.88)(.03125,-.0925){10}{\line(0,-1){.0925}}
\multiput(32.805,64.03)(.03125,-.0925){10}{\line(0,-1){.0925}}
\multiput(33.43,62.18)(.03125,-.0925){10}{\line(0,-1){.0925}}
\multiput(34.055,60.33)(.03125,-.0925){10}{\line(0,-1){.0925}}
\multiput(34.68,58.48)(.03125,-.0925){10}{\line(0,-1){.0925}}
\multiput(35.305,56.63)(.03125,-.0925){10}{\line(0,-1){.0925}}
\multiput(35.93,54.78)(.03125,-.0925){10}{\line(0,-1){.0925}}
\multiput(36.555,52.93)(.03125,-.0925){10}{\line(0,-1){.0925}}
\multiput(37.18,51.08)(.03125,-.0925){10}{\line(0,-1){.0925}}
\multiput(37.805,49.23)(.03125,-.0925){10}{\line(0,-1){.0925}}
\multiput(38.43,47.38)(.03125,-.0925){10}{\line(0,-1){.0925}}
\multiput(39.055,45.53)(.03125,-.0925){10}{\line(0,-1){.0925}}
\put(37.5,82){\vector(-1,-4){10.25}}
\put(27,80){\vector(0,-1){36}}
\thicklines
\put(33.25,87.5){\vector(0,1){.07}}\multiput(33.5,50.5)(-.03125,4.625){8}{\line(0,1){4.625}}
\put(36.75,95.75){\vector(1,3){.07}}\multiput(33.43,87.43)(.325,.825){11}{{\rule{.8pt}{.8pt}}}
\put(28.25,95.75){\vector(-2,3){.07}}\multiput(33.43,86.93)(-.47727,.79545){12}{{\rule{.8pt}{.8pt}}}
\put(33,104.75){\makebox(0,0)[cc]{Y}}
\put(73.25,104.75){\makebox(0,0)[cc]{Y}}
\put(33.75,5.5){\vector(0,1){5.5}}
\put(48.75,24){\makebox(0,0)[cc]{$u$}}
\put(64,23.75){\makebox(0,0)[cc]{$v$}}
\end{picture}

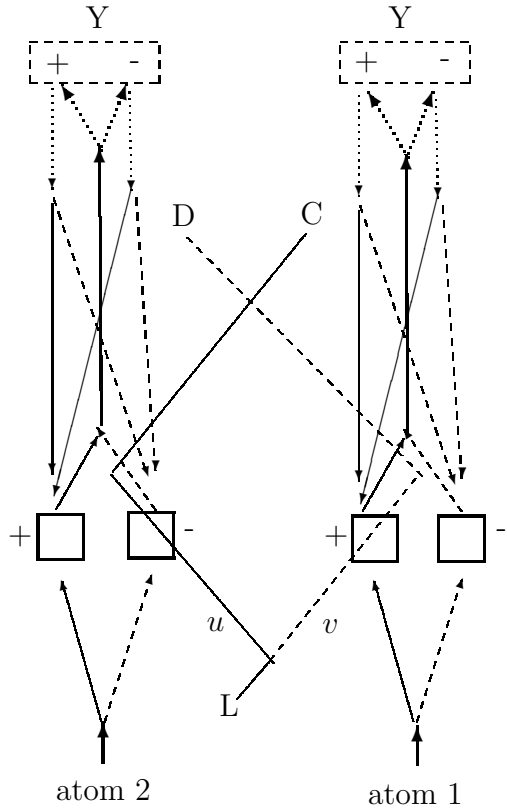
{\normalsize{Figure 4: The QLE with a later measurement of the atoms' spins along $y$. Thick lines represent the state $|x\uparrow\rangle$. Thin solid lines represent the state $|z\uparrow\rangle$; thin dashed lines represent the state $|z\downarrow\rangle$. Dotted lines represent either spin state in the $y$ direction. Arrows indicate whether these are offer or confirmation waves, with increasing time in the upward direction on the diagram.}}
\vskip .5cm

Elitzur and Dolev [2006] have argued that this situation presents a challenge for TI, for the following reason. It seems that the first transactional opportunities are for the photon OW to encounter either atom OW and to be absorbed by one or the other atom. (These transactions correspond to the four terms we omitted from the state (5) ). In the single-atom case we simply account for this with a hierarchy: if the transaction corresponding to the term $\ketv |z\uparrow\rangle$ fails, then the photon continues on through the apparatus to other possible transactions. But in this two-atom version, absorption can fail for the corresponding terms $\ketv \bsu$ and $\ketv \ud$ (and their counterparts on path $u$) and yet others terms containing $\ketu$ and $\ketv$ still remain in play for transactions involving superpositions, such as for absorption at D in a superposition of $\ketu \bsu$ and $\ketv \bsd$. So it might seem that if transactions involving absorption of the photon by either atom fail, the photon OW has to ``try again'' and head back out to see if a superposition is available, i.e., through a transaction involving a superposition of states $\bsu$ and $\bsd$. This sounds rather cumbersome and not in keeping with TI's usual elegant account.\footnote{\normalsize {The situation in the QLE is different
from the pseudotime ``echoing'' process discussed in Cramer (1986). The latter is thought of as taking place upon receipt of a particular OW component by a particular absorber and leading to either the success or failure of a particular transaction. But once that transaction has failed (i.e. in this case the one corresponding to the photon being on path $v$), that property of the quantum system is `out of the picture.' In the case of the QLE, that transaction can fail but we still need a transaction which includes the property ``photon on path v'' to activate detectors C or D. So we have to suppose (if we wish to stick to the picture of a photon OW ``really'' traveling along path v in space time) that a whole new photon OW is generated, which skips the atom and keeps going. This is what is referred to here as ``go back home and try again''--it cannot be considered part of an echoing process, nor can it be accounted for by a hierarchy--in which a transaction, once failed, takes a particular quantum system property out of the picture.}}

 We can avoid this ``go back home and try again tomorrow'' behavior of the photon OW by expanding our conception of the ontological domain of transactions to one that corresponds to the Hilbert space of the combined system. In this account, we have to consider the entire system OW together, so that the photon OW doesn't just propagate along in ordinary physical space and encounter an atom OW in the same physical space. This ``larger'' space (or a-spatiotemporal realm)
 permits different levels of transactions: those which project out a portion of the total system (i.e., absorption of the photon at the location of atom 1 or atom 2), or those which encompass the entire system (e.g., superpositions of the photon/two-atom states).

 Thus the QLE forces us to face head-on the issue sometimes raised in objections to TI: how can the OWs and CWs be ``physically present in space''(as Cramer says in his (1986))  in situations involving more than one particle? It is suggested here that, rather than consider this a problem for TI, we should consider it an opportunity. Clearly, when we consider experiments like the QLE in the usual conceptual way, we encounter nothing but paradoxes and contradictions, which are always the hallmark of a constraining paradigm. (Although many investigators of quantum theory claim that they are not hindered by unacknowledged constraining classical notions such as the idea that photons are persistent corpuscles following determinate trajectories even in the absence of measurement, the continued discussions in the literature around terms like ``which-way information,'' and arguments about whether or not a given post-selection measurement can be taken as revealing ``which slit'' a photon went through, indicates that the determinate corpuscle/trajectory paradigm is still very much with us. Furthermore, although the present discussion is
 restricted to nonrelativistic quantum theory interpretations, it should be noted that Fraser (2008) has
 presented persuasive arguments against the notion of persistent particles in the context of quantum field
 theory. Bohmians are, of course, immune to the former criticism but probably not to the latter one.) We can break through the impasse by viewing offer and confirmation waves not as ordinary waves in spacetime but rather as ``waves of possibility'' that have access to a larger physically real space of possibilities. So let us call this version of TI ``Possibilist TI" or PTI.\footnote{\normalsize {I owe this suggestion to an anonymous referee.}}

 Another advantage of this picture is its harmonious accomodation of the idea of ``quantum wholeness,'' the idea that systems defined locally are properly viewed as at least potentially dependent on their context, including phenomena physically located beyond the range of a light signal. This ontology assumes that all offer and confirmation waves should be considered as product states of all the particles in the universe, therefore having access to the entire range of transactional possibilities present in that space. Normally we don't need to consider that whole space because we are only dealing with experiments involving a tiny subset of the universal system (e.g., 1 or 2 particles in a simple product state with the rest of the universe), and transactions only project out phenomena involving the one or two particle(s). (Thus, the account given in part 2, using language such as ``the photon CW component propagates back along path $u$,'' needs to be understood as a kind of shorthand dealing only with a projection of the total system CW which actually resides in a larger space. Obviously, the question of how ordinary spacetime ``fits'' into the purported a-spatiotemporal realm proposed here is a matter for ongoing investigation; the present proposal should be viewed only as a starting point for those investigations.) When we get into more complex experiments with larger numbers of entangled particles, the available transactions become more numerous and complex as well. But the point is that the wholeness implicitly exists, at least in the form of a product space of all emitted offer waves, until a particular transaction event is realized.\footnote{ {\normalsize This in turn suggests that particle creation (understood as field quanta, not classical corpuscles) corresponds to the creation of possibility.}}

                                  \vskip .5cm
{\bf 4. The Hanbury-Twiss Effect in the QLE}   \vskip .5cm \vskip .5cm

As noted earlier, several authors{\footnote{\normalsize{E.g., EDZ (2002) and Silberstein, Stuckey and Cifone (2008). But it should be noted that these authors seem to fall short of providing a full ``relational blockworld'' (RBW) account of the QLE in that paper as claimed, since the entanglement of the atoms is not explained purely via a spacetime symmetry formulation, but is instead assumed by way of the standard quantum state for the combined dynamical system, the components of which (i.e., photons and atoms) are viewed as not ontologically fundamental in the RBW view.}} note that the EPR correlations obtaining between the atoms when detector D clicks can be achieved by using two spatially
 separated photon sources. This is seen as creating a situation where the EPR correlations are brought about not
 by any event in the atoms' past but by an event in their future (the detection of the photon at D). The analysis offered here indicates
 that the correlations are created by the joint effect of the indeterminacy of the photon's emission point and the detection at D, so there is a pre- and post-selection aspect. Under TI, the same features would seem to be present in the 2-laser case; i.e., this would be seen as functionally  equivalent to a photon from a single laser being split at S1. So transactionally, the two versions of the QLE are essentially
 equivalent in the functioning of the subsystems (photon and atoms) involved. We can see this by reviewing Cramer's account of the H-T effect.

   As Cramer explains in his 1986 paper, under TI, ``particles transferred have no separate identity apart from the satisfaction of [the] quantum mechanical
 boundary conditions.'' (1987, 678). In the H-T effect, each source (in this case an astronomical object) emits OWs that travel to the two detectors (used to make the relevant measurements). Each detector receives a
 composite OW and responds with a matching CW. For incoherent sources, the probability of a transaction is very small but still
 possible. Whether one uses coherent or incoherent sources, the effect is the same: there is no localisable ``fact of the matter'' about the photon's
 origin, since the photon is just a transaction satisfying boundary conditions. As Cramer notes of the H-T effect: ``...neither of the photons detected can be said to have originated uniquely in one of the
 two sources. Each detected photon originated partly in each of the
 two sources. It might be said that each source produced two
 half-photons and that fractions from two sources
 combined at a detector to make a full-size photon.''\footnote{\normalsize {It should be noted that coherency is not an absolute requirement for the QLE, as pointed out in Elitzur \& Dolev, 2005, p. 344, footnote 12.}}

\vskip .5cm

 {\bf 5. Conclusion}   \vskip .5cm

The Transactional Interpretation has been applied to various interaction-free measurements (IFM), including the ``quantum liar experiment'' (QLE). It has been argued that TI continues to provide an elegant and natural account of both observed quantum phenomena and of essential but heretofore mysterious features of the theory--especially of the Born Rule calling for the squaring of wave function amplitudes to obtain empirical content--provided that we consider offer and confirmation waves as residing in a ``higher'' or external ontological realm corresponding to the Hilbert space of all quantum systems involved. This a-spatiotemporal realm can be considered as a physically real space of possibilities; thus ``real'' is not equivalent to ``actual.' Therefore, this proposal can be characterized as a version of possibilist realism constrained by specific physical law. It should be emphasized that the ``possibilist realism'' proposed here is not something assumed {\it a priori}, but which emerges as a natural consequence of ``listening to the formalism'' and seeing what it may be trying to tell us about reality.

 In a way, this approach is not completely new, since
Heisenberg talked about quantum mechanics as seeming to require a notion of ``potentia,'' wherein only one outcome is experienced out of
many possible ones indicated by the formalism:\vskip .2cm

``The probability wave of Bohr, Kramers, Slater...was a quantitative version of the old concept of ``potentia" in Aristotelian philosophy. It introduced something standing in the middle between the idea of an event and the actual event, a strange kind of physical reality just in the middle between possibility and reality. (Heisenberg (2007), p. 15)\vskip .2cm

This proposal merely chooses to take seriously Heisenberg's concept of potentia and follow
it where it leads. The advantages are: \vskip .2cm(1) the Born Rule is transparently explained as a direct result of a real physical
quantity, the amplitude of the returning CW at the emitter. \vskip .2cm (2) the formalism of quantum theory, including its explicit time-symmetric aspect, is accomodated in a natural way. \vskip .2cm (3)
 it provides further insight into the origins of ``quantum wholeness.'' In this picture, actualized phenomena constitute just the ``tip of the iceberg'' of a space of mutually interacting (by way of their overall quantum state), physically real possibilities.

 Furthermore, this proposal seems harmonious with Cramer's ``transactional paradigm of time'' (2005), in which he discusses the ``emergence of present reality from future possibility'' and, in an engaging metaphor, compares this process
 to frost forming on a windowpane:
 \vskip .3cm\normalsize

 ``In the transactional interpretation, the freezing of possibility into reality as the future becomes the present is not a plane at all, but a fractal-like surface that stitches back and forth between past and present, between present and future...the emergence of the unique present....is rather like the progressive formation of frost crystals on a cold windowpane. As the frost pattern expands, there is no clear freeze-line, but rather a moving boundary, with fingers of frost reaching out well beyond the general trend...in the same way, the emergence of the present involves a lacework of connections with the future and the past, insuring that the conservation laws are respected...'' (2005, 9)
 \vskip .3cm

\large The ontology proposed herein can be seen as a perhaps unexpected but necessary feature of the ``stitching'' of the present from possibilities, by acknowledging the structured set of all those possibilities as ``too big'' to fit into ordinary spacetime. Indeed, one could even suppose that the realization of these transactional possibilities helps to create spacetime as a kind of epiphenomenon, as in the picture proposed by ED (2005, 346):
 \vskip .3cm\normalsize
 ``The wave function evolves beyond the `now', i.e., outside of spacetime, and its `collapse,' due to the interaction with the other wave functions, creates not only the events but also the spacetime within which they are located in relation to one another.''
 \vskip .3cm\large
 However, I should note that in
 this version of TI the wave function is not taken as fundamental, since neither particles nor spacetime are considered
 fundamental; the wave function, being the projection of the Hilbert space vector on the position
 basis and residing in a mathematical configuration space, implicitly assumes a substantival view of spacetime and implies
 a persistent particle ontology which is rejected in this approach.
 Rather, offer waves are
 considered to be represented by kets in Hilbert space and confirmation waves by their duals (bras)
 in a dual Hilbert space. Hilbert space and its dual are thus taken together as the basic mathematical
 representation of the dynamical domain of possibilities
 and their interactions. The wave function can
 be a useful tool in that it allows us to associate a possible spacetime location with an observed completed transaction,
 but it should not be considered as describing an amplitude of a real wave in a physically real configuration space,
 in contrast to the Bohm theory.

 Regarding the usual objection to the idea of reifying supposedly ``unphysical'' entities such as multi-system
 state vectors, it should be noted that
 when it was first discovered that there were half-integral solutions to the eigenvalue
 equation for the angular momentum operator, these might well have been considered unphysical.
 Indeed there is no ordinary spatial way to understand
 half-integer spin values--in particular the spin one-half value for the electron, which is supposedly a point particle. This serious conceptual problem is dealt with by calling electron spin an ``internal'' angular momentum, with the vast majority of phyicists not batting
 an eye over the obvious  conceptual inconsistency of a point particle having an ``internal'' degree of freedom.
  It seems to this author that the decision to take as physically real an {\it ``internal'' angular momentum of a particle that has no interior} is no less problematic than taking offer and confirmation waves as physically real.\footnote{\normalsize {And see footnote 3 for a suggestions as to how TI may provide an elegant explanation for spin as well.}}  The former assumes there is some hidden and inaccessible ``internal space'' that is nevertheless real, while the latter assumes that there is some ``larger'' and inaccessible space (corresponding to the dimensions of the relevant Hilbert space) that is nevertheless real. If one is willing to consider the
  ``internal space'' of electron spin as physically real, then it seems inconsistent to rule out the ontology proposed here on the grounds
  that it does not coincide with ordinary spacetime.

  Interestingly, one could see this further elaboration of Cramer's original 1986 interpretation as consistent with a quotation he
  provides therein from C. F. von Weizsaecker: ``What is observed certainly exists; about what is not observed we are still free to make suitable assumptions. We use that freedom to avoid paradoxes.'' (1986, 650)

Finally, to return to the QLE:
under TI, at no time was one or the other atom ``really'' in the way of a photon traveling as a corpuscular blob down an
  arm of the MZI, so we avoid the paradoxes and contradictions
  that arise in the usual pseudo-classical account of the QLE
  as in the above description of the Schr\"{o}dinger cat's schizophrenic
  history and the E-D `quantum liar' linguistic
 trap. TI interprets ``particles'' as field quanta, not as little classical
 corpuscles moving along trajectories. Thus, under TI the EPR state is brought about through the interactions of offer and
 confirmation  waves which make possible several distinct
 transactions. When one of those is actualized, discrete transfers
 of energy and other physically detectable quantities occur, which
 we like to characterize as ``particles.''

 Perhaps the most intriguing aspect of the QLE is this: that the boundary conditions for a D transaction can be satisfied by having the atoms in a superposition of ``silent detectors'', as long as each term in the superposition allows a silent detector on one arm and a vacant box on the other. It is these superposed states of the combined system that come into transactional play when the atoms' spins are measured along a direction other than $z$. Clearly we cannot think of the atoms as physically confined in the boxes in such cases, nor can the combined system's state vector 'fit' into ordinary spacetime; all of which (in the author's view) underscores the value of a quantum ontology involving a larger physical space.

\vskip .5cm
 Acknowledgements

 The author gratefully acknowledges valuable comments by Avshalom Elitzur and Joseph Kahr, a critical reading of an earlier version of this paper by John G. Cramer, and helpful comments and criticisms from an anonymous referee. (No endorsement of the views expressed herein is implied, and any errors are the author's responsibility.)
        \newpage

 References\vskip .5cm

 Berkovitz, J. (2002). ``On Causal Loops in the Quantum Realm,'' in T. Placek and J. Butterfield (Ed.], {\it Proceedings of the NATO Advanced
Research Workshop on Modality, Probability and Bell's Theorems}, Kluwer, 233-255.

 Cramer J. G. (1980). ``Generalized absorber theory and the Einstein-Podolsky-Rosen paradox.'' Phys. Rev. D 22, 362-376.

 Cramer J. G. (1986). ``The Transactional Interpretation of Quantum Mechanics.'' {\it Reviews of Modern Physics 58}, 647-688.

 Cramer J. G. (2005).
 ``The Quantum Handshake: A Review
of the Transactional Interpretation of Quantum Mechanics,'' presented at
 ``Time-Symmetry in Quantum Mechanics'' Conference, Sydney,
Australia, July 23, 2005. Available at: http://faculty.washington.edu/jcramer/PowerPoint/Sydney\_20050723\_a.ppt

Cramer J. G. (2006). ``A transactional analysis of quantum interaction-free measurements,''  {\it Found. Phys. Lett. 19}, 63-73.

 Cramer J. G. (2006). ``The Plane of the Present and the New Transactional Paradigm of Time,''  http://arxiv.org/ftp/quant-ph/papers/0507/0507089.pdf

 Elitzur A. C., Dolev S. (2006). ``Multiple Interaction-Free Measurements as a Challenge to the Transactional Interpretation of Quantum Mechanics,'' in Sheehan, D. [Ed.] {\it Frontiers of Time: Retrocausation - Experiment and Theory.} AIP Conference Proceedings 863}, 27-44.

Elitzur A. C., Dolev S. (2005). ``Quantum Phenomena Within a New Theory of Time,'' in {\it Quo Vadis Quantum Mechanics?}, Elitzur, Dolev and Kolenda, eds., Berlin: Springer, 325-349.

 Elitzur A. C., Dolev S. and Zeilinger A.. (2002) ``Time-Reversed EPR and the Choice of Histories in Quantum Mechanics,'' in {\it Proceedings of XXII Solvay Conference in Physics, Special Issue, Quantum Computers and Computing}, 452-461.

 Elitzur A. C. and Vaidman L. (1993). ``Quantum mechanical interaction-free measurements,'' {\it Found. Phys. 23}, 987-97.

 Fraser D. (2008). ``The fate of particles in quantum field theories with interactions,'' {\it Stud. His. Phil. Mod. Phys.} 39, 841-859.

 Hardy L. (1992). ``On the existence of empty waves in quantum theory.'' {\it Phys. Lett. A 167}, 11-16.

 Heisenberg (2007). {\it Physics and Philosophy}, Harper Perennial Modern Classics edition.

 Kastner, R. E. (2006). ``Cramer's Transactional Interpretationa and Causal Loop Problems.'' {\it Synthese 150}, 1-14.

 Maudlin, T. (2002). {\it Quantum Nonlocality and Relativity: Metaphysical Intimations of Modern Physics.} (Second Edition) Wiley-Blackwell.

 Schr\"{o}dinger, E. (1931). ``Über die kräftefreie Bewegung in der relativistischen Quantenmechanik'' (``On the free movement in relativistic quantum mechanics"), {\it Berliner Ber.}, pp. 418-428 (1930); ``Zur Quantendynamik des Elektrons,'' {\it Berliner Ber.}, pp. 63-72.

 Silberstein M., Stuckey W. M. and  Cifone M. (2008). ``Why quantum mechanics favors adynamical and acausal interpretations such as relational blockworld over backwardly causal and time-symmetric rivals,'' {\it Stud. Hist. Phil. Mod. Phys. 39}, 736-751.

 \end{document}